\documentclass[runningheads]{llncs}
\usepackage[T1]{fontenc}
\usepackage{graphicx,verbatim}
\usepackage{amsmath}
\usepackage{algorithm}
\usepackage{algorithmic}
\usepackage{booktabs}
\usepackage{multirow}
\usepackage{graphicx}
\begin{document}
%
% \title{IDStain: Image Disentanglement Using Color Quantization and Structure Restaining for Stain Normalization}
\title{StainPIDR: A Pathological Image Decoupling and Reconstruction Method for Stain Normalization Based on Color Vector Quantization and Structure Restaining}
%
\begin{comment}  %% Removed for anonymized MICCAI 2025 submission
\author{First Author\inst{1}\orcidID{0000-1111-2222-3333} \and
Second Author\inst{2,3}\orcidID{1111-2222-3333-4444} \and
Third Author\inst{3}\orcidID{2222--3333-4444-5555}}
%
\authorrunning{F. Author et al.}
\institute{Princeton University, Princeton NJ 08544, USA \and
Springer Heidelberg, Tiergartenstr. 17, 69121 Heidelberg, Germany
\email{lncs@springer.com}\\
\url{http://www.springer.com/gp/computer-science/lncs} \and
ABC Institute, Rupert-Karls-University Heidelberg, Heidelberg, Germany\\
\email{\{abc,lncs\}@uni-heidelberg.de}}

\end{comment}

\author{Zheng Chen}  %% Added for anonymized MICCAI 2025 submission
\authorrunning{Zheng Chen}
\institute{School of Computer Science, Wuhan University \\
    \email{2019302100067@whu.edu.cn}}

\maketitle              % typeset the header of the contribution
\begin{abstract}
The color appearance of a pathological image is highly related to the imaging protocols, the proportion of different dyes, and the scanning devices. Computer-aided diagnostic systems may deteriorate when facing these color-variant pathological images. In this work, we propose a stain normalization method called StainPIDR. We try to eliminate this color discrepancy by decoupling the image into structure features and vector-quantized color features, restaining the structure features with the target color features, and decoding the stained structure features to normalized pathological images. We assume that color features decoupled by different images with the same color should be exactly the same. Under this assumption, we train a fixed color vector codebook to which the decoupled color features will map. In the restaining part, we utilize the cross-attention mechanism to efficiently stain the structure features. As the target color (decoupled from a selected template image) will also affect the performance of stain normalization, we further design a template image selection algorithm to select a template from a given dataset. In our extensive experiments, we validate the effectiveness of StainPIDR and the template image selection algorithm. All the results show that our method can perform well in the stain normalization task. The code of StainPIDR will be publicly available later.
% Authors must provide keywords and are not allowed to remove this Keyword section.

\end{abstract}

\section{Introduction}
Pathological images stained by different dyes or scanned by different scanners often have variations in color appearance. These variations can be handled well by pathologists but this is not the case with computer-aided diagnostic systems which may encounter performance deterioration on images with different color domains \cite{van2021deep,srinidhi2021deep,verghese2023computational}. So it is vital to normalize stained images before using them for model training and lots of research has been carried out to handle this problem.

Traditional stain normalization methods without deep learning algorithms often follow the idea of factorizing images into a color matrix and a structure matrix \cite{macenko2009method,gavrilovic2013blind,vahadane2016structure} by certain matrix factorization algorithm. Then, traditional methods complete the normalization procedure by simply multiplying the structure matrices with a target color matrix (factorized from a template image). These kinds of methods can generate images with a highly consistent color appearance but as the decomposing and combining procedures have a relatively high information loss, the generated images sometimes can hardly improve the performance of the downstream tasks \cite{hetz2024multi}.

Deep-learning-based stain normalization methods mainly adapt the generative models to pathological image datasets \cite{shaban2019staingan,salehi2020pix2pix,zhou2019enhanced,kablan2024stainswin,hetz2024multi,kang2021stainnet}. These methods all encode the pathological images to latent features and decode them to generate normalized images. Due to the larger model parameters and more calculations, deep-learning-based methods often outperform traditional methods in the stain normalization task. But as the transform procedure is all completed by the decoder, they implicitly transfer the images to the target domain which sometimes leads to inconsistency in normalized images. 

To combine the strength of consistent color appearance in traditional methods and effectiveness in deep-learning methods, we propose an image decoupling and reconstruction algorithm that follows and expands the traditional idea. We first decouple a pathological image into a structure feature and a vector-quantized color feature. Then by restaining the structure feature with a target vector-quantized color feature and decoding the stained structure feature, the proposed method explicitly performs the task of stain normalization. In contrast to previous traditional methods which need to manually select a template image to get the target color feature, we additionally proposed a template image selection method to choose a suitable template image in a given pathological image dataset.

The main contribution of this paper can be summarized as follows:
\begin{itemize}
    \item [$\bullet$] A novel image decoupling and reconstruction architecture is proposed to conduct the stain normalization task.
    \item [$\bullet$] An insight about color discreteness and finiteness in the single color domain which inspired us to discretize color features by the codebook.
    \item [$\bullet$] An effective stain module based on mixed cross-attention is proposed to fuse color/structure features in latent space.
    \item [$\bullet$] A novel template image selection method is proposed to solve the manual selection problem of template images in previous works.
\end{itemize}

\section{Method}
\subsection{StainPIDR Overview}
\begin{figure}[h]
\centering
\includegraphics[width=1\textwidth]{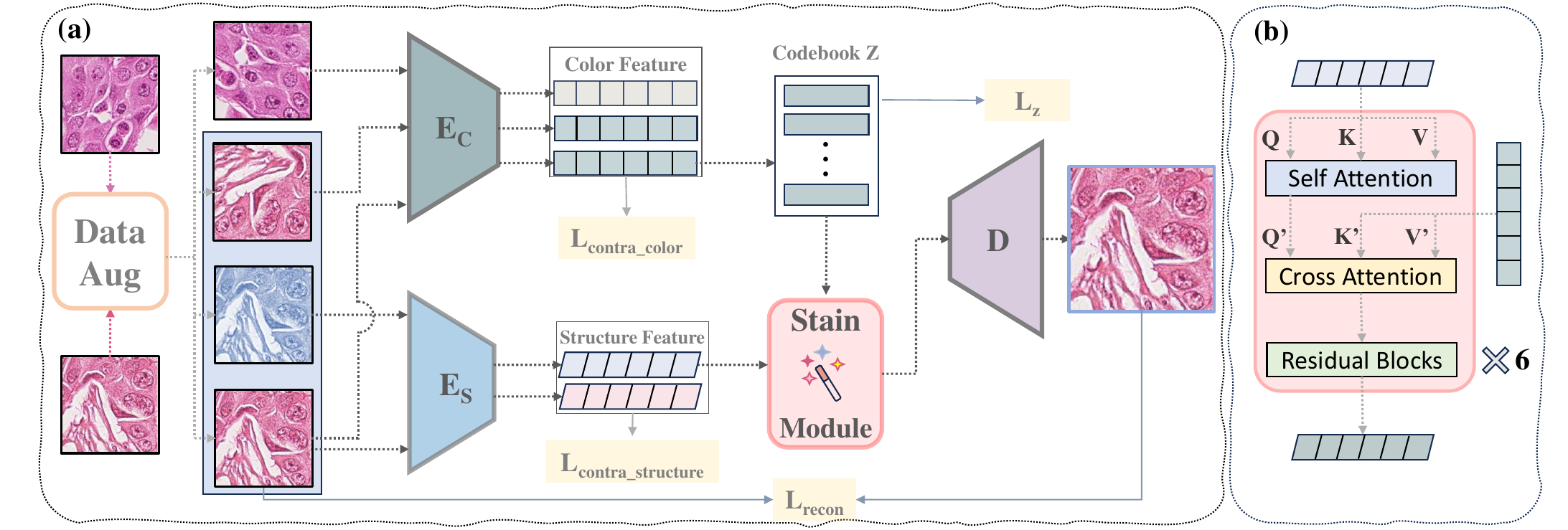}
\caption{Overview of our StainPIDR method. (a) is the training procedure of our model. The pathological image will first be augmented to put into $\mathrm{E_C}$ and $\mathrm{E_S}$, the output vector of $\mathrm{E_C}$ will be quantized and put into SM with the output vector of $\mathrm{E_S}$ together, and finally, the decoder D will generate an image based on the output of the stain module. (b) shows the architecture of our stain module.}
\label{algorithm_overview}
\end{figure}

Our pathological image decoupling and reconstruction method follows the basic idea of traditional methods of decoupling a simple pathological image into a structure feature and a color feature. As we assume that decoupled color features from one color domain should be discrete and finite, we further discretize the vectors of color features. Then, we restain the structure features with the target vector-quantized color features by the cross-attention mechanism and decode the restained structure features to normalized pathological images. The StainPIDR can finally generate desirable pathological images with excellent structure preservation. Figure \ref{algorithm_overview} (a) shows the overview of our StainPIDR method. There are five components in total: the color encoder($\mathrm{E_C}$), structure encoder($\mathrm{E_S}$), color vector quantization codebook(Z), stain module(SM), and image decoder(D). At the inference stage, we further propose a template image selection algorithm to get the template image for color decoupling.

\subsection{Pathological Image Decoupling and Reconstruction}

To efficiently extract the color attribute and the structure attribute from a pathological image, we send two images from two different color domains into our model. The two images will first be augmented to train the $\mathrm{E_C}$ and $\mathrm{E_S}$ encoders as depicted in Figure \ref{algorithm_overview} (a). For $\mathrm{E_C}$ and $\mathrm{E_S}$, we utilized the contrastive learning idea for training. For $\mathrm{E_C}$, there are three input images: two color-identical images (A and $\mathrm{A'}$) and one image of a different color (B) and three outputs: $\mathrm{E_{C}(A)}$, $\mathrm{E_{C}(A')}$, and $\mathrm{E_{C}(B)}$. The contrastive loss of $\mathrm{E_C}$ contains the maximum similarity of $\mathrm{E_{C}(A)}$ and  $\mathrm{E_{C}(A')}$ and the minimum similarity of $\mathrm{E_{C}(A)}$ and $\mathrm{E_{C}(B)}$.

\begin{small}
\begin{equation}
\begin{aligned}
L_{contra}(E_C, A, A', B)=1 - CosineSimilarity(E_C(A), E_C(A')) +\\ 
 1 + CosineSimilarity(E_C(A), E_C(B))
\end{aligned}
\end{equation}
\end{small}

For $\mathrm{E_S}$, there are two input images and we adopted the same strategy as $\mathrm{E_C}$ but with no minimum similarity part. To get better performance, we adapted the model architecture in ConvNext \cite{liu2022convnet} to extract the features.

\begin{small}
\begin{equation}
\begin{aligned}
L_{contra}(E_S, A, A'')=1 - CosineSimilarity(E_S(A), E_S(A''))
\end{aligned}
\end{equation}
\end{small}

After that, we applied a codebook $\mathit{Z}$ to map the vectors of color features $E_C(A)$ to the most similar vectors in the codebook under the assumption that decoupled color features from one color domain should be discrete and finite. This vector quantization operator ${Z}(z)$ can be formulated as:

\begin{small}
\begin{equation}
\begin{aligned}
{Z}(z):=\mathop{argmin}\limits_{z_i\in{Z}}|| z - z_i ||
\end{aligned}
\end{equation}
\end{small}

Let $sg[\cdot]$ be the stop gradient operator. The embedding loss for training the codebook can be formulated as:

\begin{small}
\begin{equation}
\begin{aligned}
L_{Z}(Z, E_C, A)=|| sg[E_C(A)]-Z(E_C(A)) || + \alpha|| sg[Z(E_C(A))] - E_C(A) ||
\end{aligned}
\end{equation}
\end{small}

Then, we put $Z(E_C(A))$ along with the output of $E_S$ into the stain module, as shown in Figure \ref{algorithm_overview} (b). This stain module contains 6 stain blocks with each consisting of a self-attention layer for feature sharpening, a cross-attention layer for feature fusion, and a residual layer for feature retaining.

Finally, we put the output of the stain module into the decoder D to reconstruct the pathological image. Accordingly, we have a reconstruction loss when evaluating the quality of generated images.

\begin{small}
\begin{equation}
\begin{aligned}
L_{recon}(E_C, E_S, Z, SM, D, A)=E_{a \sim p_{data }(a)}[|| D(SM(E_S(a), Z(E_C(a)))) - a ||_2]
\end{aligned}
\end{equation}
\end{small}

Through the above procedure, the trained StainPIDR model can efficiently decouple an image into structure features and vector-quantized color features. By restaining the structure features with specified color features and decoding, the StainPIDR method can generate normalized images with good structure preservation.

\subsection{Template Image Selection Algorithm}

At the inference stage, there is a problem with how to select a template image from a given pathological image dataset. Traditional methods like Reinhard, Macenko, and Vadahane need to manually select a template image for normalization. However, this manual selection is subjective, and hard to guarantee a good performance. So we proposed an algorithm $S^*$ to select a template image from the image dataset which can have a relatively good performance in downstream tasks. 

\begin{small}
\begin{equation}
\begin{aligned}
S^*(A):=\mathop{argmin}\limits_{a_i\in{A}}WD(E_{a \sim p_{data }(a)}[Histogram(a)], Histogram(a_i))
\end{aligned}
\end{equation}
\end{small}

As the main cause of performance decrease lies in the color appearance, we try to quantify the color of an image by using its color histogram. Accordingly, we calculate the Wasserstein Distance \cite{ramdas2017wasserstein} (WD) on color histograms between different images to measure their color distance. Following this thought, we first calculate the mean color histogram of the whole dataset which is reckoned as the color representation of the whole pathological image dataset. Then, we calculate the WD between the color representation and each image. The image with a minimum distance from the color representation is selected as our template image to perform the stain normalization task.

% \renewcommand{\algorithmicrequire}{\textbf{Input:}}
% \renewcommand{\algorithmicensure}{\textbf{Output:}}
% \begin{algorithm}
%     \caption{Target Image Selection $S^*$}
%     \begin{algorithmic}
%         \ENSURE{Selected Target Image from Dataset $i^*$}
%         \REQUIRE{Pathological Image Dataset with Target Stains $D=\{i_1, i_2,..., i_n\}$}
%         \STATE{average_histogram = }
%         \FOR{image number $k \in [1, n]$}
%             \IF{$x\leq0$}
%                 \STATE{$y\gets-1$}
%             \ELSE
%                 \STATE{$y\gets1$}
%             \ENDIF
%         \ENDFOR
%     \end{algorithmic}
% \end{algorithm}

\section{Experiments}

\subsection{Experimental Setup}
To validate our StainPIDR method, we conducted four experiments on three datasets: MITOS12 \cite{ludovic2013mitosis}, MITOS14, and GlaS \cite{sirinukunwattana2017gland}. Besides, we conducted two ablation experiments to study the effect of different modules and the template image selection algorithm. In MITOS12 and MITOS14, there are pathological images scanned by two different scanners: Aperio Scanscope and Hamamatsu. All the pathological images in these three datasets were cropped to 256 $\times$ 256. For MITOS12 and GlaS datasets, we retained the excess pixels by cropping from the end. For MITOS14, we simply discarded the excess pixels. The ground truth label files in MITOS12 and annotation masks in GlaS were processed accordingly. All the processed images were categorized as train datasets and test datasets with two color domains in MITOS12 and MITOS14.

Among the four experiments, there are two image quality assessment experiments and two downstream experiments. For the image quality assessment experiments, we normalized pathological images from Hamamatsu-scanned to Aperio-like on MITOS12 and normalized in a different direction in MITOS14. We evaluated image quality by three metrics: SSIM \cite{assessment2004error}, MS-SSIM \cite{wang2003multiscale}, and UQI \cite{wang2002universal}. The two downstream experiments are a mitosis detection experiment and a gland segmentation experiment. For the mitosis detection task, we used a YOLOv5 model to detect mitosis on the unnormalized images and normalized images with mAP50 and mAP50-95 to evaluate. Each stain normalization method was also trained on MITOS12 datasets to normalize images in MITOS12 from Aperio-scanned to Aperio-like. For the gland segmentation task, we used a Unet model to do the segmentation and evaluated each method by Dice, Accuracy, Precision, and Recall metrics. Each stain normalization method was trained on MITOS14 datasets and we normalized images in GlaS to Aperio-like.

In the ablation experiments, we removed the color vector quantization codebook and the stain module to study their effect on the StainPIDR. To validate the template image selection algorithm, we further randomly select ten images from MITOS12 test datasets scanned by Aperio. We conducted the same mitosis detection experiment on MITOS12 with all these ten template images.

We compared StainPIDR with previous state-of-the-art methods with publicly available codes including Reinhard\cite{reinhard2001color}, Macenko\cite{macenko2009method}, Vahadane \cite{vahadane2016structure}, StainGAN \cite{shaban2019staingan}, MultiStain-CycleGAN\cite{hetz2024multi} (MCN for simplicity).

\subsection{Implementation Details}

For the training settings of our StainPIDR, we use the AdamW optimizer and set the learning rate, beta 1, beta 2 to 1.5e-4, 0.9, 0.95. We trained our model for 300 epochs in the MITOS12 dataset and 200 epochs in the MITSO14 dataset.

For all the traditional methods, we use the implementation in StainTools\footnote{https://github.com/Peter554/StainTools}. For all the other compared models in our experiment, we trained them as the original settings in their open-source code repository.
% https://github.com/Peter554/StainTools
% https://github.com/twpkevin06222/Gland-Segmentation
% https://github.com/ultralytics/yolov5
For the Unet baseline model in gland segmentation, we adopted the implementation\footnote{https://github.com/twpkevin06222/Gland-Segmentation} in GitHub. We removed the color augmentation and didn't load the pre-trained weights. We trained this Unet model for 200 epochs. For the YOLOv5 baseline model in mitosis detection, we adopted the implementation\footnote{https://github.com/ultralytics/yolov5} in GitHub. We removed the color augmentation and trained the YOLOv5 model for 100 epochs.

\subsection{Results}

In this section, we will present the results of all four experiments. First, we show the qualitative results on the MITOS14 and MITOS12 datasets.

\begin{figure}[h]
\centering
\footnotesize
\includegraphics[width=0.7\textwidth]{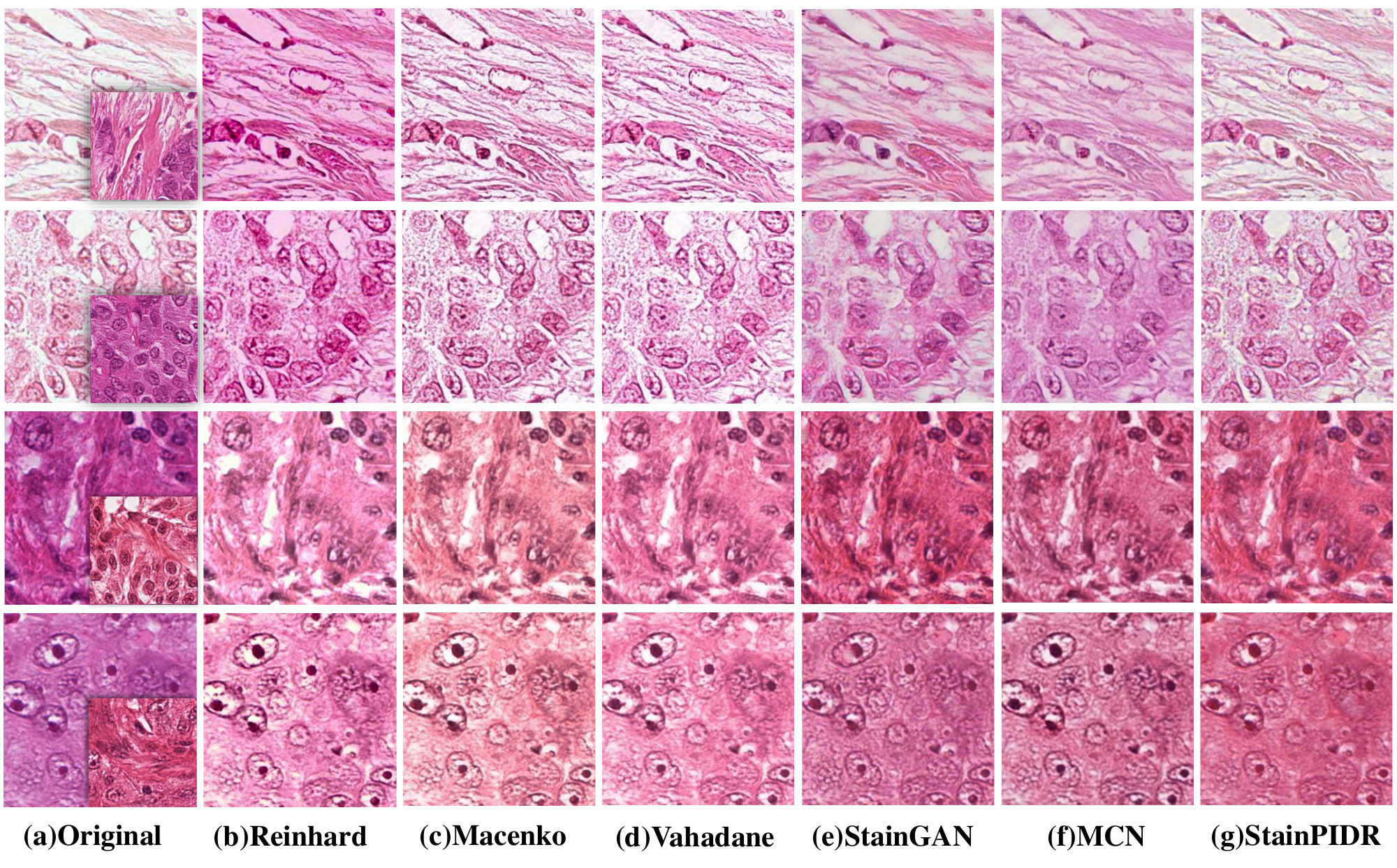}
\caption{Qualitative results on MITOS14 and MITOS12 datasets. The first two rows show the normalized images of MITOS14 and the third and fourth rows show the normalized images of MITOS12.}
\label{mitos14_results}
\end{figure}

Figure \ref{mitos14_results}(b)-(g) shows some of the normalized images on the two datasets. Figure \ref{mitos14_results} (a) shows the original images. The attached images on them are the target images scanned by another device and the images attached in the first and third rows are used as the templates for traditional methods and ours. We can find that images normalized by Reinhard, StainGAN, and MCN (MultiStain-CycleGAN) have an undesirable background in MITOS14. Reinhard makes the background stained. StainGAN and MCN make the background a little bit gray. Besides, images normalized by StainGAN still have a tiny color difference. Macenko sometimes normalizes images to an undesirable color.

\begin{table}
\caption{Evaluation scores on MITOS14 for SSIM, MS-SSIM, and UQI}
\label{mitos14_table}
\centering
\scriptsize
\begin{tabular}{llccc|ccc}
    \toprule
    & & \multicolumn{3}{c}{MITOS14} & \multicolumn{3}{c}{MITOS12} \\
    \midrule
    & methods & SSIM $\uparrow$ & MS-SSIM $\uparrow$ & UQI $\uparrow$ & SSIM $\uparrow$ & MS-SSIM $\uparrow$ & UQI $\uparrow$\\
    \midrule
    & Reinhard \cite{reinhard2001color}     &0.957&0.981&0.973&0.924 &0.966 &0.954\\
    & Macenko \cite{macenko2009method}      &0.932&0.975&0.977&0.919&0.949&0.951 \\
    & Vahadane \cite{vahadane2016structure} &0.918&0.967&0.973&0.921&0.960&0.955 \\
    & StainGAN \cite{shaban2019staingan}    &0.832&0.939&0.983&0.924&0.949&\textbf{0.986} \\
    & RestainNet \cite{zhao2022restainnet}  &0.916&0.976&0.973&/&/ &/ \\
    & RRAGAN \cite{baykal2023regional}      & 0.933 & 0.965 & 0.987&/&/ &/ \\
    & StainSWIN \cite{kablan2024stainswin}  & 0.943 & 0.976 & 0.991&/&/ &/ \\
    & MCN \cite{hetz2024multi}              &0.852&0.910&0.983&0.918&0.961&0.981 \\
    & StainPIDR                                  & \textbf{0.974} & \textbf{0.991} & \textbf{0.993}&\textbf{0.974}&\textbf{0.974}&0.981\\
    \bottomrule                         
\end{tabular}
\end{table}

To quantitatively measure the structure preservation ability of each method, we further calculated their scores on three metrics: SSIM, MS-SSIM, and UQI. Table \ref{mitos14_table} shows the quantitative results of each method. There are three more results because these methods all conducted the same experiment on the MITOS14 dataset but didn't open their source code. So we directly added their results to our table. From the table, our StainPIDR method achieves the best structure preservation ability among all the others.

Although our method achieved the highest structure preservation scores, downstream tasks are the most effective way to evaluate a stain normalization method. We further conduct two downstream tasks to confirm its validity. The first downstream task is a gland segmentation task on the GlaS dataset. We directly used the trained stain normalization models on MITOS14 for the normalization of images in GlaS. We trained and tested the Unet model on each normalized dataset. The second downstream task is a mitosis detection task on the MITOS12 dataset. We trained stain normalization models on it and applied the trained model to normalize the images. Table \ref{downstream_tasks_results} shows our results. 

\begin{table}
\caption{Performance of each method on downstream tasks}
\label{downstream_tasks_results}
\centering
\scriptsize
\begin{tabular}[width=0.6\textwidth]{lcccc|cc}
    \toprule
     & \multicolumn{4}{c}{Gland Segmentation} & \multicolumn{2}{c}{Mitosis Detection} \\
    \midrule 
        Images & \makebox[0.1\textwidth][c]{Dice$\uparrow$} & \makebox[0.1\textwidth][c]{Acc$\uparrow$} & \makebox[0.1\textwidth][c]{Pre$\uparrow$} & \makebox[0.1\textwidth][c]{Recall$\uparrow$} & \makebox[0.1\textwidth][c]{$\mathrm{mAP_{50}}$$\uparrow$} & $\mathrm{mAP_{50-95}}$$\uparrow$\\
    \midrule
    Original                              & 0.850 & 0.890 & 0.894 & 0.897 & 0.714 & 0.471\\
    Reinhard \cite{reinhard2001color}     & 0.852 & 0.895 & \textbf{0.912} & 0.882 & 0.704 & 0.477\\
    Macenko \cite{macenko2009method}      & 0.849 & 0.893 & 0.901 & 0.893 & 0.679 & 0.440\\
    Vahadane \cite{vahadane2016structure} & 0.854 & \textbf{0.900} & 0.906 & 0.900 & 0.715 & 0.468\\
    StainGAN \cite{shaban2019staingan}    & 0.848 & 0.889 & 0.886 & \textbf{0.903} & 0.715 & 0.484\\
    MCN \cite{hetz2024multi}              & 0.777 & 0.828 & 0.796 & 0.900 & 0.684 & 0.455\\
    StainPIDR                                  & \textbf{0.858} & \textbf{0.900} & 0.907 & 0.899 & \textbf{0.735} & \textbf{0.494}\\
    \bottomrule                         
\end{tabular}
\end{table}

This result shows that the stain normalization of pathological images can improve the performance of downstream tasks. But sometimes, the normalized images may also decrease the model performance. The reduction of MCN may be caused by its grayscale of images which have a certain information loss. By contrast, our proposed StainPIDR method can still perform well in both downstream tasks.

\begin{figure}[h]
\centering
\includegraphics[width=1\textwidth]{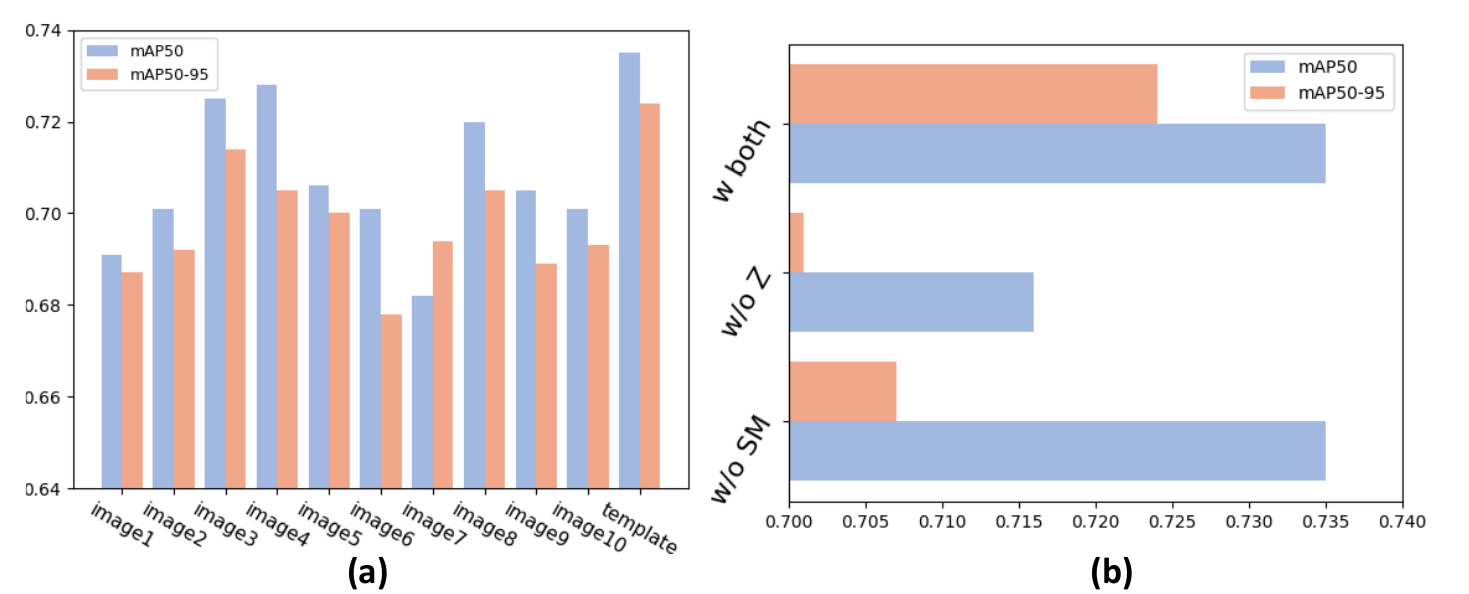}
\caption{Ablation studies. (a) validation of template selection algorithm. (b) validation of the color vector quantization codebook (Z) and the stain module (SM). All the mAP50-95 results added 0.23.}
\label{ablation_results}
\end{figure}

Figure \ref{ablation_results} (a) shows our ablation results. In Figure \ref{ablation_results} (a) images 1 to 10 are the randomly selected images and the template is the image selected by our selection algorithm from the MITOS12 test dataset. Our template selection algorithm has the highest performance among all these images, which validates the effectiveness of the proposed algorithm. In Figure \ref{ablation_results} (b) we showed the different influences of each module in StainPIDR. The removal of the stain module or the color vector quantization codebook both decreases the performance, which confirms our suggestion of color discreteness and the effectiveness of cross attention.

\section{Conclusion}

In this paper, we proposed a novel image decoupling and reconstruction method called StainPIDR for the stain normalization task. Our StainPIDR method is illuminated by the traditional stain normalization methods and expands them. We discretize the color features and restain the structure features by cross-attention. Further, we propose a template image selection algorithm to find a good template image from the given dataset. From our experiments, it is convincing that normalized images by StainPIDR can effectively improve the performance of downstream tasks. However, in the inter-domain normalization tasks, StainPIDR still needs to be enhanced. In the future, we will continue to enhance our method for inter-domain normalization tasks.

%splncs04
\bibliographystyle{elsarticle-num} 
\bibliography{citations}

\end{document}